\documentclass[manuscript]{acmart}
\AtBeginDocument{%
  }

\setcopyright{acmlicensed}
\copyrightyear{2026}
\acmYear{2026}
\acmDOI{10.1145/3773077.3812170}

\copyrightyear{2026}
\acmYear{2026}
\setcopyright{cc}
\setcctype{by}
\acmConference[IDC '26]{Proceedings of the 25th Interaction Design and Children Conference}{June 22--25, 2026}{Brighton, United Kingdom}
\acmBooktitle{Proceedings of the 25th Interaction Design and Children Conference (IDC '26), June 22--25, 2026, Brighton, United Kingdom}
\acmDOI{10.1145/3773077.3812170}
\acmISBN{979-8-4007-2283-7/2026/06}

\begin{document}

\title[Emergent Technology, Emergent Critique: Students and Teachers Developing Critical AI Literacy]{Emergent Technology, Emergent Critique: Students and Teachers Developing Critical AI Literacy through Participatory Design around Generative AI}

\author{Santiago Ojeda-Ramirez}
\email{sojeda@uci.edu}
\affiliation{%
  \institution{University of California, Irvine}
  \city{Irvine}
  \state{California}
  \country{USA}
}

\author{Eva Durall Gazulla}
\email{eva.durallgazulla@oulu.fi}
\affiliation{%
  \institution{University of Oulu}
  \city{Oulu}
  \country{Finland}
}

\author{Kylie Peppler}
\email{kpeppler@uci.edu}
\affiliation{%
  \institution{University of California, Irvine}
  \city{Irvine}
  \state{California}
  \country{USA}
}

\renewcommand{\shortauthors}{Ojeda-Ramirez et al.}

\begin{abstract}
Who gets to decide how generative AI tools enter students' classrooms?
We report on a five-week participatory design program in which three
11th-grade Latinx students and three high school teachers in California
negotiated how generative AI tools would be used and taught about in
learning environments. Drawing on video recordings and designed
artifacts, we ask: what critical AI literacy practices emerged as
students and teachers jointly designed how generative AI tools would be
used and taught about? Our analysis reveals three practices:
collectively unsettling assumptions about AI, mutual learning through
complementary expertise, and grounding AI critique in cultural knowledge
and creative practice. Students and teachers developed these practices
through the design work itself. This case contributes strategies for
designing with youth around an emergent technology like generative AI
toward critical AI literacy. It extends work on youth as protagonists by
showing how this approach enables students to shape both the adoption
and the interrogation of these tools in their learning environments.
\end{abstract}

\ccsdesc[500]{Human-centered computing~HCI design and evaluation methods}
\ccsdesc[300]{Social and professional topics~K-12 education}

\keywords{critical AI literacy, participatory design, co-design, generative AI, high school, Latinx youth, AI literacy}

\maketitle

\section{Introduction}

Generative AI in education has the hallmarks of an emerging technology
\cite{rotolo2015emerging}: it is growing fast, its impact on learning
is widely anticipated, and yet which tools enter classrooms, how, and
toward what ends remains genuinely uncertain. As Frauenberger et al.\
\cite{frauenberger2025emerging} argue, when a technology is emergent,
the stakeholders affected by the adoption of such technologies have a
democratic claim to participate in shaping them. Thus, participatory
design (PD) is essential to ensure that these systems are designed and
deployed in ways that remain accountable to the needs and practices of
those most directly impacted.

A growing body of research has worked to make AI legible to young
people, helping them understand how these systems work and how to build
with them
\cite{han2025empowering,dangol2025childrens,long2023fostering}, with
researchers asking not just whether youth can learn AI, but whether
they can engage with it as questioners of its design and use:
interrogating whose data trains models, who bears the costs of
algorithmic bias, and how technical systems encode societal inequities
\cite{veldhuis2025critical,aleman2024data,kenny2025beyond}. Prior
studies have used participatory approaches to design how AI literacy
gets taught \cite{noh2026youcanactually,mawasi2023learning}, but the
question of who shapes which aspects of AI are worth questioning, and
from whose social and community position, has gone largely unasked.
Critical AI literacy \cite{veldhuis2025critical} is precisely the
orientation needed to ask it: not just how AI tools enter learning
environments, but what about them deserves scrutiny, whose knowledge
frames that scrutiny, and what it means to teach AI critically. In this
paper, we examine what it produces when high school students bring that
orientation as design protagonists, shaping how generative AI tools are
used and taught about in their own learning environments.

We conducted a five-week PD program with three 11th-grade students and
three high school teachers in a southern California city, in which each
teacher-student dyad produced a curricular unit shaping how generative
AI tools would be used and taught about in their subject area. Our
descriptive case study addresses the following question: What critical
AI literacy practices emerged as high school students and teachers
jointly designed how generative AI tools would be used and taught about
in their learning environments? Three practices emerged: (1)
collectively unsettling assumptions about AI; (2) mutual learning
through complementary expertise; and (3) grounding critique in cultural
knowledge and creative practice. This paper contributes a situated case study with two interrelated
claims: first, it offers strategies for PD around an emergent
technology like generative AI, oriented toward fostering critical AI
literacy practices; and second, it extends work on youth as
protagonists
\cite{iversen2017child,mahboob2021re,weixelbraun2024discussing} by
showing what that positioning requires in formal education contexts,
where generative AI is still emergent and teachers are co-designers. We
show that when students hold genuine design authority in that setting,
they shape not only how these tools are adopted but also what about them
becomes an object of scrutiny and whose knowledge determines what gets
taught.

\section{Background}

\subsection{Critical AI Literacy as a Situated Practice}

AI literacy has been conceptualized as a set of competencies enabling
individuals to critically evaluate AI technologies, communicate and
collaborate effectively with AI, and use AI as a tool in everyday life
and work \cite{long2020ai,yim2024critical}. Calls within computing
education have pushed beyond competency frameworks to ask whose
interests are served by what we teach about technology, and to make the
ethical and political dimensions of computing available to all students
\cite{vakil2019power}. Critical AI literacy extends this foundation by
foregrounding the sociopolitical dimensions that conventional frameworks
tend to underemphasize. Veldhuis et al.\ \cite{veldhuis2025critical}
synthesize critical AI literacy across four interrelated dimensions:
disrupting the commonplace, or questioning taken-for-granted assumptions
about AI in everyday life; considering multiple viewpoints, including
those of marginalized communities; focusing on the sociopolitical, by
analyzing how AI is entangled with power, inequality, and institutional
structures; and taking action, by translating critical understandings
into ethical and design-oriented interventions.

These dimensions describe orientations and objectives, leaving open the
social processes through which they develop. Drawing on Lave and
Wenger's \cite{lave1991situated} conceptualization of learning as
situated participation in social practice, we understand critical AI
literacy as something that emerges \textit{through the doing}, through
negotiations, decisions, and collective sense-making among participants.
Aleman and Martinez \cite{aleman2024data,aleman2025directions} showed
that community situatedness was essential to how youth developed
critical orientations toward AI; similar work in southern California
communities has demonstrated how learners' cultural resources shape
their engagement with AI concepts \cite{ojedaramirez2024ailiteracy}.
Logan \cite{logan2024learning} argued that learning about generative AI
requires mapping its political-economic ecologies; empirical work by
Durall et al.\ and Babai et al.\ has further documented how teenagers
engage in ethical reflection when AI activities are embedded in school
contexts \cite{durall2025youth,babai2025navigating}. Together, these
studies suggest that critical AI literacy emerges through participation
in particular kinds of social activity
\cite{iivari2023computational,hartikainen2023proud,iivari2024fostering}.
We use Veldhuis et al.'s \cite{veldhuis2025critical} four dimensions as
our guiding lens.

\subsection{Shaping AI Literacy Learning Environments Through Participatory Design}

AI literacy frameworks center students as mindful users of AI
\cite{long2020ai}. Positioning them instead as designers of AI uptake
\cite{mahboob2021re} raises deeper questions: which tools enter their
learning environments, what about them deserves scrutiny, and whose
knowledge shapes what gets learned. CCI research has long been
concerned with positioning youth as active designers of computing
technologies
\cite{druin2002role,kafai1995minds,iversen2017child,iivari2018empowering}.
Iversen, Smith, and Dindler \cite{iversen2018computational} proposed
computational empowerment as a framework for how youth engage critically
with technology through construction and deconstruction activities
\cite{iversen2017child}. Iivari and colleagues
\cite{iivari2022critical,iivari2024transformative} have argued that
truly empowering designs must embed critical and justice-oriented
orientations from the outset, and a broader program of research has
examined what conditions sustain that empowerment
\cite{vanmechelen2021systematic,smith2024agenda,dindler2023dorit,schaper2023empowerment,shokeen2025sure}.

This critical turn within CCI has begun to shape how researchers
approach AI literacy learning. Several studies have engaged youth as
designers of AI systems, showing that critical and ethical
understandings can emerge through the construction process itself
\cite{moralesnavarro2025babygpts,moralesnavarro2025ijcci,moralesnavarro2024peerauditors,moralesnavarro2022codesigning,kafai2021codequilt,weixelbraun2024discussing}.
Others have involved teachers as designers in PD processes around AI
literacy, demonstrating that PD with educators increases agency and
supports the integration of critical perspectives on AI into learning
environments \cite{noh2026you,bilstrup2025automation}.
Mawasi et al.\ \cite{mawasi2023learning} showed that when youth and
educators design together, relations of expertise are redistributed and
both parties learn from each other in ways neither anticipated. When PD
has included students, they have been positioned as designers of AI
systems or as advisors to teacher-led curriculum design
\cite{noh2025youth,cai2025childai}, leaving open the question of what
it looks like when students shape the critical AI literacy learning
space through PD. This is the gap our study addresses.

\section{Methods}

\subsection{Participants and Context}

We conducted an after-school PD program with three 11th-grade students
and three high school teachers at a public high school in a southern
California city, where 98\% of students identify as Latinx. The
students, Klaus, Sunny, and Violeta (two girls and one boy), were
enrolled in courses taught by their teacher-collaborators: Duncan, a
computer science teacher; Isa, a design teacher; and Le\'{o}n, a social
studies teacher. All names are pseudonyms.

\subsection{Workshop Activities}

The program consisted of five weekly co-design sessions of 90 minutes
each, held between February and April. Each session followed a
consistent three-part structure. Sessions opened with a
critical-ethical icebreaker activity in which participants collectively
discussed a sociotechnical dilemma related to AI, such as algorithmic
bias in college admissions, environmental costs of data centers, or
authorship in AI-generated art. These openings were designed to
position AI as a public, political, and cultural problem from the
outset, anchoring the co-design work in critical AI literacy before
design work on the units began. The middle portion of each session
combined short conceptual inputs with hands-on co-design work in
teacher-student dyads, where each teacher partnered with a student to
engage in backward lesson planning, activity design, or assessment
design. Sessions closed with whole-group discussion or presentations,
surfacing insights and tensions across the dyads and fostering shared
reflection on how AI, ethics, and community intersected in their
designs.

Sessions 1 and 2 established conceptual foundations: acknowledging
different forms of AI literacy, and introducing critical and speculative
design \cite{dunneraby2013} as frameworks, with participants drafting
learning objectives, outcomes, and assessment ideas. Each teacher
entered the program with a vision for how to integrate generative AI
tools into their subject area; students and teachers together shaped
the terms of that integration, producing a curricular unit in which
specific tools would be engaged, questioned, and taught critically.
Sessions 3, 4, and 5 are the focus of this paper. Across them, dyads
moved from conceptual discussion toward designing concrete activity
sequences, drawing on generative AI tools (including ChatGPT, Claude,
and Gemini) and evaluating them for bias in outputs and relevance to
their students' community and lives.

\subsection{Data Collection and Analysis}

We collected four primary sources of data: (1) video and audio
recordings of all five PD sessions, (2) co-designed curriculum
artifacts produced across the program including activity sequences,
unit plans, and design worksheets, (3) pre and post interviews with all
six participants, and (4) researcher memos written after each session.
We constructed a descriptive case study following Yin \cite{yin2018case},
as enunciated by Morales-Navarro et al.\
\cite{moralesnavarro2025babygpts} in the context of CCI research: a
detailed description of a phenomenon in context, aimed at depicting how
this particular group of participants engaged in PD across a sustained
program of work, attending to particulars over causal explanation. The
first author coded the video recordings of Sessions 3, 4, and 5, which
are the focus of this paper. We organized the video logs using Veldhuis
et al.'s \cite{veldhuis2025critical} four dimensions as an organizing
lens, supplemented with co-designed artifacts and interview excerpts.

\section{Findings}

Three practices of critical AI literacy emerged across the focal
co-design sessions, each reflecting a different mode of collective
engagement with AI's sociotechnical dimensions. Across all three, what
made the practices possible was not only that students and teachers
worked together, but that the PD structure made each participant's
distinct knowledge necessary. Each dyad inflected these practices through their subject area: the
science fiction unit foregrounded sociopolitical futures, and the
design unit centered community conditions and justice.

\subsection{Collectively Unsettling Assumptions about AI}

Each session opened with an ethical icebreaker that brought AI's real
consequences into the room before any curriculum design work began, and
that lived experience became material participants could design with.

In Session~3, Klaus and Le\'{o}n were working on backwards planning for
a science fiction unit. As they discussed how to structure students'
introduction to AI, Klaus made a design proposal rooted in what the
group had just lived through: a series of small structured discussions
around the dilemmas described above (e.g., algorithmic bias in college
admissions, environmental costs of data centers, or authorship in
AI-generated art), what he called `baby fishbowls,' a structured
discussion format in which a small group debates a scenario while
others observe, supporting active listening and meaningful participation
across the group. Students would grapple with those scenarios and use
that as the launch for the entire unit. His reasoning was explicit:
``if you didn't know anything about AI, you probably didn't also know
that it used up that much water\ldots{} I think that by having these
conversations, they would also learn about all these consequences.''
The icebreaker format was now something Klaus understood as a
pedagogical structure worth encoding into the unit.

Le\'{o}n recognized the pedagogical value immediately, noting the
fishbowl format ``coincides with the stuff admin is pushing right
now.'' Klaus's proposal arrived with a theory drawn from lived
experience: students learn to question AI by confronting its real
consequences in a structured space for disagreement. That theory was
available to Klaus precisely because the participatory design structure
asked him to carry two roles simultaneously: student who had just had
an assumption unsettled, and designer deciding how to unsettle
assumptions for others.

This practice reflects Veldhuis et al.'s \cite{veldhuis2025critical}
dimension of disrupting the commonplace: making strange what students
might otherwise take for granted about AI. That disruption emerged
through the PD task itself. The backwards planning activity required
Klaus to answer what encountering these tools should feel like, and the
answer he produced was shaped by having already questioned them:
structured confrontation with what AI does and costs, before any
unexamined use begins.

\subsection{Mutual Learning through Complementary Expertise}

The teacher-student dyad structure created conditions in which
expertise flowed bidirectionally, a dynamic PD scholars describe as
mutual learning \cite{robertson2014mutual}. Designing a unit for the
teacher's own learning environment through PD meant student knowledge
was structurally necessary: the question of what to teach about AI
could not be settled without knowing what it meant to be a student
encountering AI in this community. What this session shows is that in
the context of critical AI literacy, that complementarity is not
incidental; the student's community knowledge was what shaped the
critical interrogation of AI.

In Session~2, Isa and Violeta were working on a restaurant unit's
activity sequence. Isa turned the design question directly to Violeta:
``maybe you can help me plan that. How they use a chatbot for the
restaurant industry. How would that be used?'' She was asking her
design partner to contribute knowledge she needed. Violeta's answer
functioned as a design argument: ``I would imagine, as they use it for
customer service. Because you can't really replace a lot of things in
the restaurant with AI, because it's all about the people that make
it.'' She proposed specific forms: chat boxes at each station, AI
handling dietary adjustments and loyalty programs. The reasoning came
from knowing what restaurants in a predominantly Latinx community are
and what would be lost if the people in them were displaced. Isa built
directly on Violeta's framing and it shaped the activity sequence
forward.

This exchange illustrates Veldhuis et al.'s \cite{veldhuis2025critical}
dimensions of considering multiple viewpoints and focusing on the
sociopolitical. Isa brought pedagogical expertise; Violeta brought
knowledge of what hospitality means in this community. The dyad made
both forms necessary, and the unit carried the shape of both.
Crucially, Violeta's contributions were constitutive, not
supplementary, to the design work. Her positioning as a designer is
what gave her knowledge, rooted in questioning what these tools would
displace and for whom, the authority to determine how AI would figure
in that learning environment and what about it would be taught.

\subsection{Grounding AI Critique in Cultural Knowledge and Creative Practice}

The third practice emerged from what the design tasks required: that
participants name the community conditions surrounding their students'
lives. In Session~4, Isa and Violeta worked on a restaurant unit's
activity sequence. When the researcher introduced systems thinking as a
design frame, Isa invoked a design justice poster on her classroom
wall: ``who has a seat at the table, and who is your design for? Who
is it helping?'' This was a design criterion for evaluating what AI
would do in that space. She then proposed a pedagogical pivot: after
students complete their business plans, she would ask them to redesign
for 2035, imagining what AI's role in the restaurant might become.
``We have large natural resources, yet large food insecurity,'' she
said, naming the material conditions of the community her students come
from and positioning those conditions as the actual design problem. The
restaurant had become a scenario rooted in the lived experiences of
Isa's students and their community.

\begin{figure}[h]
  \centering
  \includegraphics[width=0.9\columnwidth]{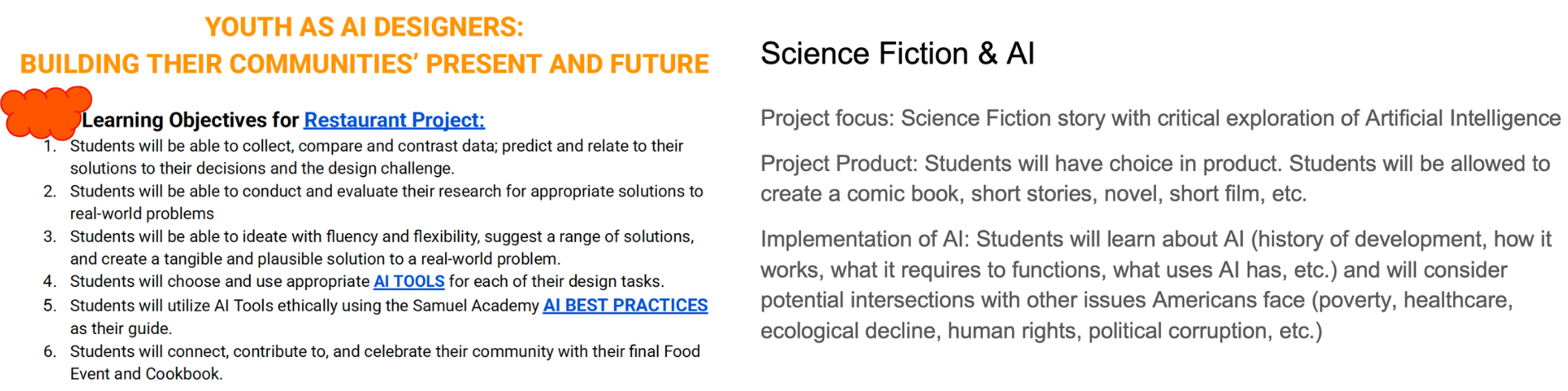}
  \caption{Two co-designed artifacts showing critical AI literacy
  practices materialized in curriculum. Left: Le\'{o}n and Klaus's
  science fiction unit. Right: Isa and Violeta's restaurant unit.}
  \Description{Two hand-written planning artifacts from the
  participatory design sessions. Left shows a science fiction unit plan
  organized around community conditions and AI futures. Right shows a
  restaurant activity sequence with AI ethics integrated before tool
  use.}
  \label{fig:artifacts}
\end{figure}

In Session~5, Le\'{o}n and Klaus presented their unit design.
Le\'{o}n described asking students to research poverty, healthcare,
ecological decline, human rights, and political corruption, then
imagine AI futures through those lenses (see
Figure~\ref{fig:artifacts}). Klaus described why the icebreaker
structure had to anchor the unit: ``the icebreakers both gave us data
and information about AI in society, and then at the same time, asked
us to use that data and information in our unit.'' The critical AI
literacy developed in the PD sessions was now being deliberately
reproduced for students who had not yet sat in those sessions. Klaus's presence across these two vignettes makes a larger arc
visible: the student who proposed the fishbowl structure in Session~3
is the same student presenting the finished unit in Session~5. This practice reflects Veldhuis et al.'s
\cite{veldhuis2025critical} dimension of taking action: the
interrogation of what AI tools do, whose knowledge they encode, and
what they cost was materialized as the content students would
encounter.

\section{Discussion}

This study asked what critical AI literacy practices emerged when high
school students and teachers jointly designed how generative AI tools
would be used and taught about in their learning environments. What the
data shows is that the practices emerged through the PD structure
itself. In each case, PD was what enabled youth and educators to
develop critical AI literacy together, determining which questions
about AI tools were worth asking, what about those tools would become
objects of scrutiny, and whose knowledge shaped those answers
\cite{dindler2020computational,iivari2024transformative,mawasi2023learning}.
Critical AI literacy was simultaneously process and outcome: students
and teachers developed it through the very act of deciding how AI tools
would be used and taught about.

What this account turns on is a distinction between students as
participatory designers and students as informants or consultants. In
studies where students contribute knowledge to adult-led design
processes, that knowledge is often treated as input to be incorporated
or set aside. Here, students' knowledge, forged through critical
questioning of what AI tools do and for whom, was constitutive,
determining how those tools would figure in their learning environments
and what about them was worth learning. That authority was available to
them because the PD structure positioned them as designers
\cite{iversen2017child,mahboob2022brave}.The dyads in this case sustained collaboration with little observable
hierarchy reassertion, a pattern worth examining across contexts. For the CCI community, this reframes participatory design as a site of
critical AI literacy practice in its own right
\cite{iivari2022critical,iivari2023computational}. It also contributes
to CCI's ongoing work on youth as protagonists
\cite{iversen2017child,mahboob2022brave} by specifying what enabled
such positioning in the context of an emergent technology in education:
the icebreaker format, the teacher-student dyad, and community-situated tasks each made student knowledge structurally necessary to shape how AI would be used and taught about. Given the small and highly contextualized sample, this case
demonstrates what is possible in this setting; future work should examine the
sustainability of these practices and how they adapt across different
school contexts. This case makes the stakes of those questions
concrete, and their answers consequential for how participatory design
with youth around generative AI in formal education is practiced.

\begin{acks}
The authors thank the students and teachers who gave their time and
expertise to this project. This work was supported by the Spencer
Foundation and the National Academy of Education as well as the
Research Council of Finland through the Critical Datalit project
\#354445.
\end{acks}

\section*{Selection and Participation of Children}

We recruited students through school and class visits to their
teachers' classrooms at a public high school in a city located in
southern California, United States. Youth participated in the study in
after-school hours. Parents received consent forms prior to the study,
which included a brief explanation of the research, and youth provided
assent to their participation. All participants received monetary
compensation for their time. Research protocols and data collection
methods were approved by the Institutional Review Board of the
University of California, Irvine.

\bibliographystyle{ACM-Reference-Format}
\bibliography{references}

\end{document}